\documentclass[amsmath,amssymb,groupedaddress,superscriptaddress,longbibliography]{elsarticle}
\usepackage{hyperref}
\usepackage{epsfig}
\usepackage{color}
\usepackage{graphicx}
\usepackage{bm}
\usepackage{epstopdf}
\usepackage{amsmath}
\usepackage{tikz}
\usepackage{todonotes}
\usepackage{subfigure}
\usepackage{textcomp}
\usepackage{wasysym} 
\usepackage{general_latex}

\begin{document}
\title{Benchmarking the multiconfigurational Hartree method by the exact wavefunction of two harmonically trapped bosons with contact interaction}  
\author{Yeongjin Gwak}
\address{Center for Theoretical Physics, Department of Physics and Astronomy, Seoul National University, 08826 Seoul, Korea}
\author{Oleksandr V. Marchukov}
\address{Technische Universit\"{a}t Darmstadt, Institut f\"{u}r Angewandte Physik, 64289 Darmstadt, Germany}
\author{Uwe R. Fischer}
\address{Center for Theoretical Physics, Department of Physics and Astronomy, Seoul National University, 08826 Seoul, Korea}

\begin{abstract}
We consider two bosons in a one-dimensional harmonic trap, interacting by a contact potential, and  compare the exact solution of this problem to a self-consistent numerical solution by using the multiconfigurational time-dependent Hartree (MCTDH) method.
We thereby benchmark the predictions of the MCTDH method with a few-body problem that has an analytical solution for 
the most commonly experimentally realized interaction potential in ultracold quantum gases.   
It is found that exact ground state energy 
and first order correlations are accurately reproduced by MCTDH up to the intermediate dimensionless coupling strengths corresponding to typical background scattering lengths of magnetically trapped ultracold dilute Bose gases. 
For larger couplings, established for example by (a combination of) 
Feshbach resonances and optical trapping, the MCTDH approach 
overestimates  the depth of the trap-induced correlation dip of 
first order correlations in position space, as well as overestimates 
the fragmentation, defined as the average relative occupation of orbitals other than the energetically lowest one. 
We anticipate that qualitatively similar features in the correlation function may arise for larger particle numbers, paving the way for a quantitative assessment of the accuracy of MCTDH by experiments with ultracold atoms.  
\end{abstract}

\date{\today}
\maketitle

\section{Introduction} 
The 
MCTDH method is a powerful self-consistent numerical approach to 
the quantum dynamics of many interacting particles, and has been 
extensively used to 
predict correlation functions, 
cf., 
e.g., Refs.~\cite{Meyer1990,meyer2009,alon2008,mctdhx,Lin2020}. 
Initially used for the purpose of propagating wavepackets in physical chemistry, where it is by now routinely used \cite{MeyerReview}, in the past decade MCTDH has increasingly been applied to describe the  intricate many-body physics of ultracold dilute Bose gases,  
for example, in Refs.~\cite{lode2012,Grond,streltsov2013,fischer2015,kroenke,BBKYMCTDH,klaiman2015,klaiman2016,sakmann2016,LodeBruder,Tsatsos2017}; see for a recent review \cite{lode2019}.  

The present study is partly inspired by the ongoing debate on the convergence of MCTDH, see, e.g., 
Refs.~\cite{DrummondBrand,SakmannReply,Olsen,cosme}. 
These convergence issues arise because the MCTDH equations of motion
become singular as soon as unoccupied orbitals occur during the real or imaginary 
time evolution. Hence, some (nonunique) prescription of 
regularization is needed, see for example \cite{manthe,KlossLubich,MeyerWang}.
In particular, the MCTDH method, based on the Dirac-Frenkel
variational principle, does not provide a rigorous control of the error of many-body evolution
such as, e.g., the McLachlan principle of least error does \cite{lee}. 
Furthermore, it is not clear 
whether MCTDH is more accurate 
in comparison to, e.g., the  alternative approach of using the truncated Wigner method 
for either large or small number of particles $N$ 
\cite{Olsen}. 
This stems from the fact that neither method, MCTDH nor truncated Wigner (see also, e.g., Ref.~\cite{DrummondWigner}) 
provides a  {\em control parameter} for its accuracy to be assessed within given numerical resources. 
This should be compared with 
(number-conserving) Bogoliubov theory 
\cite{CastinDum,Cederbaum}, where this control parameter is some power of the inverse
of the particle number, $1/N$. 
Rigorous results on the accuracy of retaining just a single orbital in the field operator expansion are available in the limit of particle number $N\rightarrow \infty$, provided the (formal)  
condition is met that the interaction coupling $g$ decreases as $1/N$, 
 and hence $g=g(N)$ tends to zero in that limit \cite{Lieb,Lieb2003}.    
These rigorous results are, in addition, limited to reproducing the Gross-Pitaevskii energy correctly, while higher-order correlations reveal deviations from mean-field physics 
even in the large $N$ limit keeping $gN$ fixed cf., e.g., \cite{klaiman2015,klaiman2016}.
 
Importantly, a direct experimental verification of the accuracy of MCTDH 
in a controllable quantum many-body system is lacking so far.
We here aim at benchmarking 
MCTDH with the exactly solvable model most closely associated with current 
experiments on ultracold gases: A pair of bosons with repulsive contact interactions trapped in a single harmonic well.
Because many-body correlations are strongest in one spatial dimension, we use to this end a one-dimensional (1D) variant of the originally 3D analytical solution 
\cite{Busch1998,Calarco}: For $N=2$ in one spatial dimension, one expects 
deviations from (single-orbital) mean-field physics to be most significant.  
The present case of strong correlations is therefore an excellent testing ground for the accuracy of MCTDH outside its usual applicability domain of weak correlations. 
Upon approaching the Tonks-Girardeau ``fermionized'' limit \cite{Minguzzi,cazalilla2011,paredes2004,kinoshita2004},  the {\em self-consistent} determination of the orbitals' shape in a harmonic trap becomes increasingly important, as the usual periodic boundary conditions in a spatially homogeneous system cannot be applied.    
While it is well known that  in 1D, the Lieb-Liniger solution 
\cite{lieb1963a} is exact for any $N$, extracting correlation functions 
is in general a challenging task \cite{cazalilla2011}. In addition,  
the Lieb-Liniger solution 
is not available in harmonic traps. 


The analytically solvable $N=2$ problem 
supplies an exact statement on the shape of the 
orbitals and level occupation statistics. 
It can thus 
assess the accuracy of MCTDH, which determines these quantities,   
for a large but finite number $M$ of field operator modes. 
We provide below, 
with an {\em experimentally realizable} interaction potential,
an accurate quantitative statement to which extent MCTDH is ``numerically exact'' \cite{sakmann2009},  i.e., 
controllably reproduces for $M\rightarrow\infty$ an exact solution of the Schr\"odinger equation 
\cite{Nnote}. 
The coupling strength 
can be changed over a large range via Feshbach resonances \cite{chin2010},
facilitating experimental access to the validity domain of MCTDH. We 
demonstrate that for large couplings, MCTDH increasingly overestimates a trap-induced 
dip in nonlocal first-order correlations, which can be used as a sensitive measure of the accuracy of MCTDH. 



\section{Analytical solution} 
The Hamiltonian is
\begin{align}
\label{hamiltonian}
 H = -\frac{\hbar^2}{2m}\Delta_{\vec x}
 + \frac{1}{2} m \omega^2  {\vec x}^2 
 + g \delta(x_1 - x_2),
\end{align}
where $\vec x =  (x_1,x_2)$ is the position vector of the atoms, $m$ their mass, $\omega$ the frequency of the trapping potential, and $g$ is the 1D interaction coupling constant.
Below, we use $\hbar \omega $ as unit of energy,  
and $l=\sqrt{\hbar/m\omega}$ as length scale. 
The solution of the Schr\"{o}dinger equation 
can be found by the separation ansatz \cite{Busch1998}  
\begin{equation}
\Psi(R, r) = \Psi_{\rm COM}(R) \psi_{\rm rel}(r),
\label{solution}
\end{equation}
where we introduced relative, $r = \frac{1}{\sqrt{2}} (x_1 - x_2)$ and center-of-mass (COM) $R = \frac{1}{\sqrt{2}} (x_1 + x_2)$ coordinates.
Relative and COM wavefunctions are then given by
\begin{align}
 \Psi_{\rm COM}(R) &  \propto e^{-R^2/2}H_n(R), \nonumber\\
 \psi_{\rm rel}(r) &  \propto  e^{-r^2/2} U(-\nu, \frac{1}{2}; r^2), 
 \label{com_and_rel}
\end{align}
where $H_n$ is the Hermite polynomial of order $n$ and $U(-a, b; x)$ is a confluent hypergeometric
function \cite{abramowitz}; we omitted the normalization constants. 
A new quantum number $\nu$ 
parametrizes the total energy  of the system 
\begin{equation}
E = 2 \nu + n + 1, 
\label{full_energy}
\end{equation}
where the $g$ dependence of $\nu$ is found by solving 
\begin{equation}
\label{eq_nu}
\frac{\Gamma(-\nu + \frac{1}{2})}{\Gamma(-\nu)} = - \frac{g}{2\sqrt{2}}.
\end{equation} 
Clearly, the wavefunction in Eq.~\eqref{solution} 
describes the system we consider exactly. 
In the following, we compare ground state energy, single-particle density matrix (SPDM) and the shape of the orbitals, obtained by employing this exact solution 
with the results from MCTDH calculations, varying the coupling $g$ and the number of orbitals $M$. We note here that the $N=2$ harmonic trap wavefunction has previously been used to compare 
to MCTDH results \cite{cosme}, however for only up to intermediate values of {\em negative} $g\sim\ord(-1)$,  for maximally $M=10$ orbitals, and without the crucial 
comparison of 
{\em nonlocal} first-order correlations we present below,  
which encapsulate MCTDH self-consistency in particular for strong correlations. 

Using Eqs.~\eqref{solution} and \eqref{com_and_rel}, the SPDM 
$\rho ^{(1)}(x, x') = \int \Psi^{\ast}(x, x_1) \Psi(x_1, x') \mathrm{d}x_1 $ 
of the ground state, which is obtained from $n = 0$ and $\nu = \nu_0$ with $\nu_0$ being the minimal value of $\nu$  from solving Eq.~\eqref{eq_nu}, is given by 
\begin{multline}
 \rho ^{(1)} (x, x') 
 \propto 
e^{ -{(x^2 + x'^2)}/{2}}
\\
\times \int  \mathrm{d}x_1 e^{-\frac{x_1^2}2} U
\left(-\nu_0, \frac{1}{2}; \frac{(x - x_1)^2}2\right) 
\times (x\leftrightarrow x').
\label{exact_dm}
\end{multline}
where the integral 
may be calculated numerically to in principle arbitrary accuracy. 
Here, $x\leftrightarrow x'$ denotes exchanging the corresponding argument  
in the preceding expression, reproducing it else identically.

\section{The MCTDH method}  
The notion of self-consistency 
embodied by MCTDH is 
that it determines the shape and time
dependence of the {\em orbitals} $\varphi_i(x,t)$
{\em self-consistently} together with their occupation distribution
$C_{\overrightarrow{N}}(t)$
in Fock space, where $\overrightarrow{N}=(N_0, N_1, \dots, N_{M-1})$ 
($\sum^{M-1}_0 N_i = N$) is the occupation vector. The coupled 
MCDTH equations of motion are 
\begin{eqnarray}
\label{mctdh_eqs}
\!\!\! i\hbar\delderiv{\f C(t)}{t} &=& \f H(t) \f C(t), \nonumber\\
\!\!\! i\hbar \delderiv{\ket{\varphi_j}}{t} &=& \hat P\left [\hat h \ket{\varphi_j} + \sum_{k,s,q,l=1}^M \rho^{-1}_{jk} \rho_{ksql}\hat W_{sl}\ket{\varphi_q}  \right]\!.
\end{eqnarray}
Here, $\f{C}(t)$ is the column vector that consists of all possible expansion coefficients $C_{\overrightarrow{N}}(t)$, 
$\f H(t)$ corresponds to the time-dependent Hamiltonian matrix in the 
basis  $\ket{\overrightarrow{N}; t}$, $\hat h$ is the single-particle Hamiltonian, 
$\hat W_{sl}=g\int\int \mathrm{d} x 
\varphi^{\ast}_s (x) 
\varphi_l(x)$, 
and $\hat P = 1 - \sum_{k'=1}^M \ket{\varphi_{k'}}\bra{\varphi_{k'}}$ is an orthogonal subspace projection operator.  
Finally, 
$\rho_{ksql}$ 
is the matrix element of the two-particle density matrix. 
To find the self-consistent solution of the above equations, we use 
MCTDH-X software package,  provided by \cite{mctdhx, Lin2020} and first implemented in \cite{lode2016,fasshauer2016}. 

\begin{figure}[t]
\centering
\includegraphics[width=0.63\textwidth]{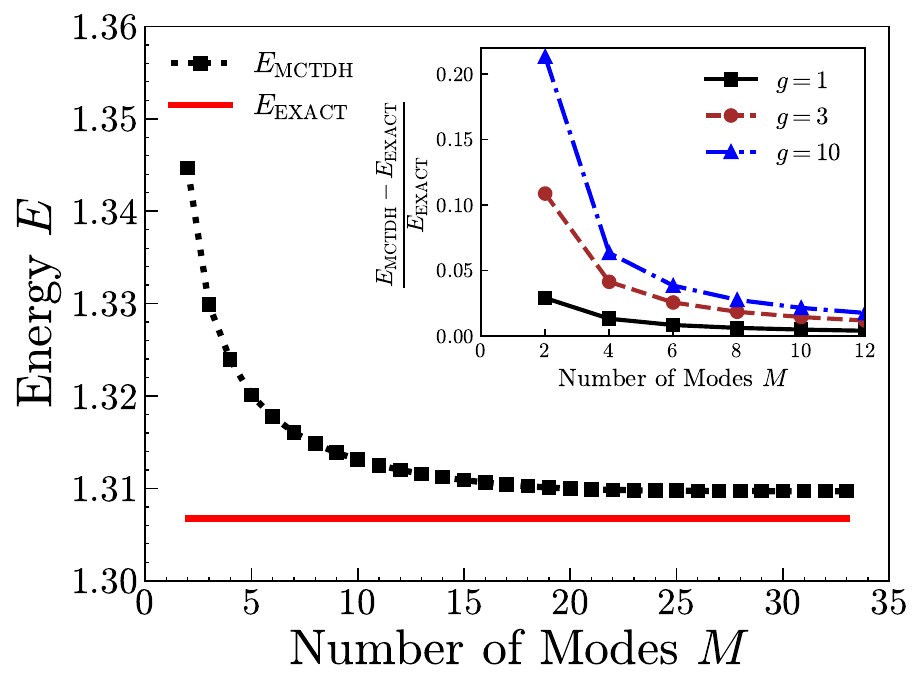}
\includegraphics[width=0.645\textwidth]{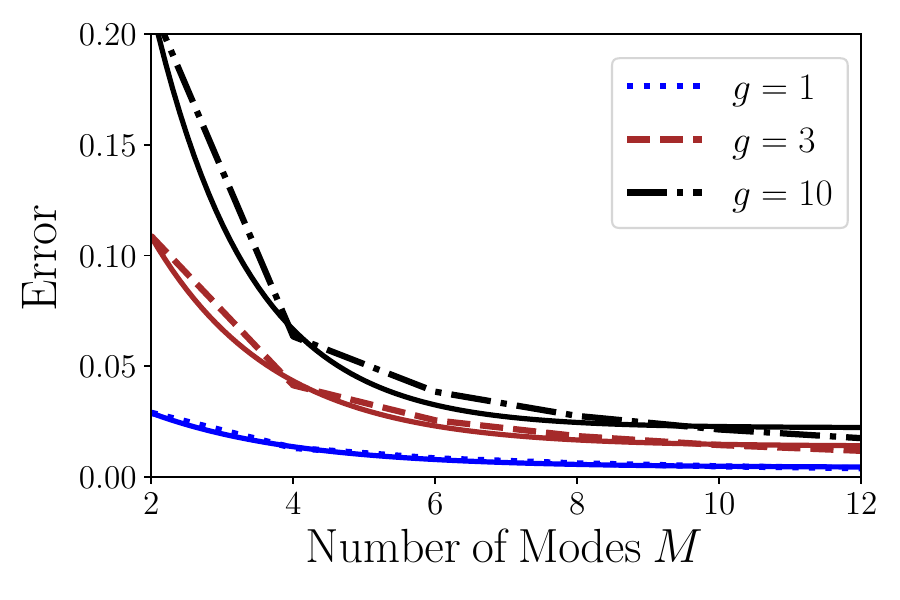}
\caption{\textit{Top:} Convergence of the ground state energy, calculated via MCTDH-X with increasing number of orbitals (black squares), $M=2,\dots,33$
towards the exact value from Eq.~\eqref{full_energy} (red solid); the coupling $g=1$.
Inset: The relative error for the ground state energy for $g=1$ (black solid), $g=3$ (brown dashed) and $g=10$ (blue dash-dotted). Exact ground energies for different values of $g$ are $E_{g=1} = 1.30675, E_{g=3} = 1.59991, E_{g=10} = 1.85069$. \textit{Bottom:} Exponential fit for the relative error for the ground state (inset of the top figure).
}
\label{fig_energy}
\end{figure}

\section{Convergence of MCTDH to exact ground state energy} 
In order to verify convergence of the ground state energy to the exact result, we performed
extensive MCTDH calculations for a wide range of the number of orbitals, $M=2,\dots, 33$. In Fig.~\ref{fig_energy}, we present the comparison between the exact and numerical values of the ground state energy  for the interaction coupling $g = 1$. We conclude that the numerical value converges rapidly for a large number of orbitals. The relative error between the exact and converged numerical values becomes less than $3$\,\textperthousand~when $M>15$. 
We however also notice that upon further increase of $M$, the error does not decrease
significantly further. Specifically, for $M=20$ the error is $2.48$\,\textperthousand, 
and for $M=33$ it is still $2.26$\,\textperthousand. 
From Fig. \ref{fig_energy} we see that for large $M$ the energy converges exponentially with a small relative error, corresponding to results of similar calculations that employed interaction rescaling, for a smaller number of orbitals, see Ref.~\cite{Ernst}. 
To illustrate the dependence of the convergence on $g$, the relative error for the energy, 
${(E_{\rm MCTDH} - E_{\rm exact})}/{E_{\rm exact}}$, is shown in the inset of  
Fig.~\ref{fig_energy} for $M=2,\dots,12$ and  $g=1, 3,$ and $10$. The bottom panel in Fig~\ref{fig_energy} shows the exponential fit of the relative error as a function of number of modes $M$ for intermediate ($g=1, 3$) and strong ($g = 10$) interaction strength, demonstrating that 
a similar behavior of the convergence is obtained for various $g$.
The MCTDH calculations still 
converge reasonably well for sufficiently large $M$ to the exact energy. However, the computational cost (the $M$ needed for convergence) is, as expected, seen to increase 
for larger values of $g$.

\section{Density matrix}  
Generally, correlation functions are more sensitive to the accuracy of MCTDH predictions 
than the ground state energy is, cf.~\cite{klaiman2015,klaiman2016,klaiman2017}. Therefore, 
we now concentrate on a comparison of the analytics to numerics in the form of the first-order correlations, as encapsulated by the SPDM. 
We compare the results of our MCTDH calculations,  
in addition, with the Monte Carlo calculations performed by Minguzzi \textit{et al.} in Ref.~\cite{Minguzzi} for the SPDM of a pair of hard-core bosons in a 1D harmonic trap.
The emphasis for this part of the paper is to assess 
the accuracy of MCTDH when $g$ in the Hamiltonian Eq.~\eqref{hamiltonian} is varied from weak over intermediate to strong coupling, so that we here fix 
$M=10$. 
\subsection{Real space}
\begin{figure}[t]
\center
\centering
\subfigure{\includegraphics[width=0.75\textwidth]{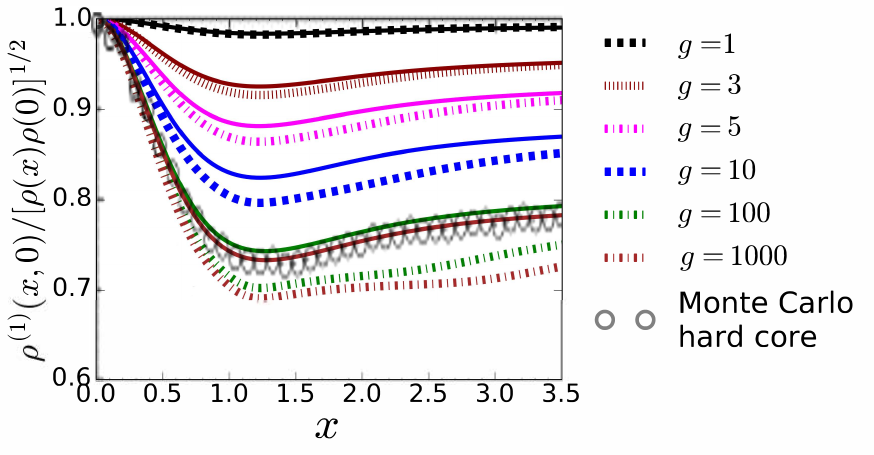}}
\hspace*{-1.5em}
\subfigure{\includegraphics[width=0.7\textwidth]{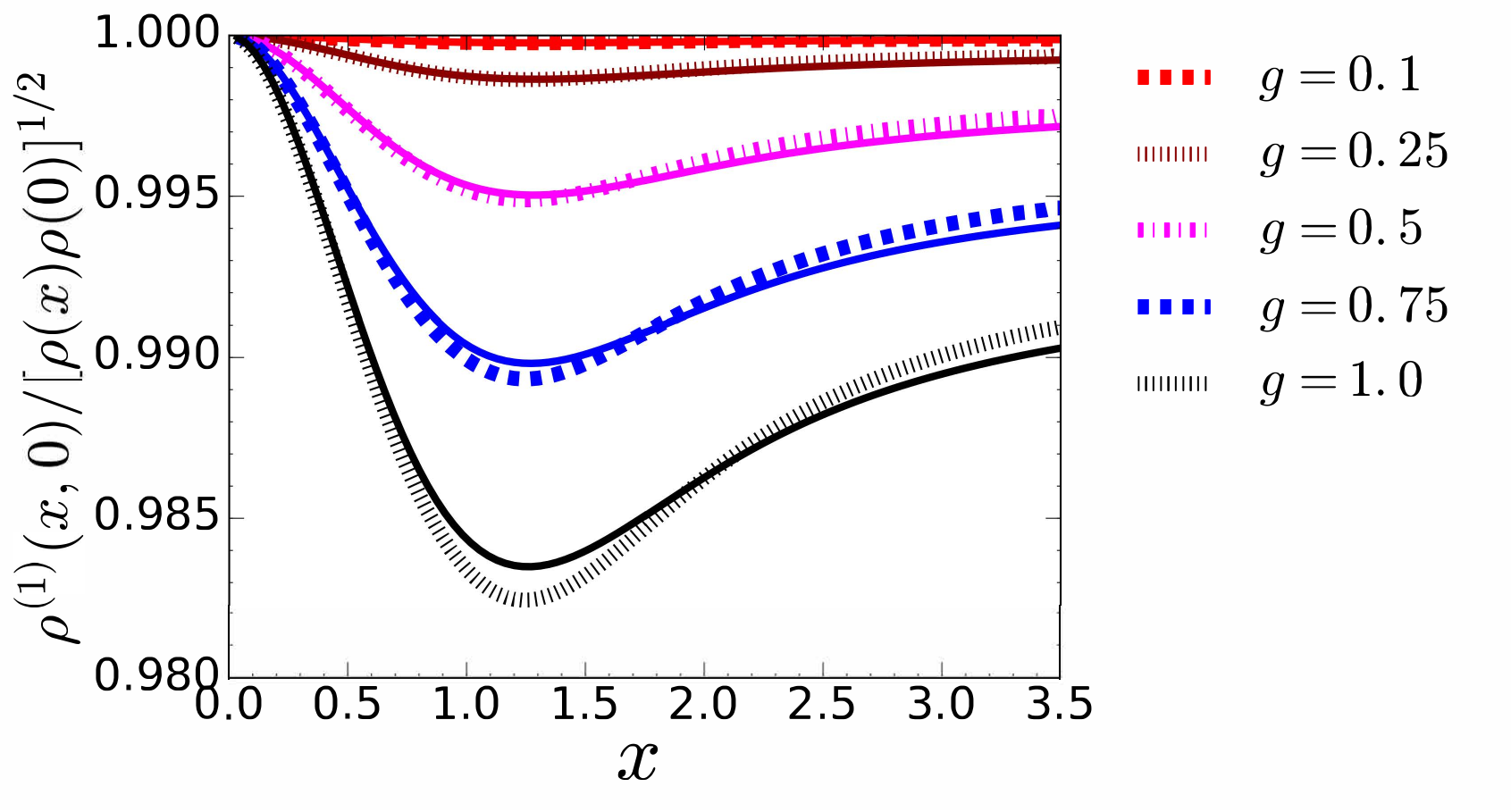}}
 \caption{Top: SPDM $\rho^{(1)}(x, x')/\sqrt{\rho(x)\rho(x')}$ as a function of $x$ and $x'=0$ for $N=2$ interacting bosons in a harmonic trap, in the strong coupling regime. The gray circles are the Monte Carlo results of Minguzzi et al.~\cite{Minguzzi} for hard-core bosons ($g\rightarrow \infty$), with the size of the circles representing the error bars in the Monte Carlo data. The lines are MCTDH results for various  
 $g$ and $M=10$. The solid lines show the analytical result. 
Bottom: Comparison of MCDTH results ($M=10$) 
with the 1D analytical solution 
in the range of intermediate interaction couplings.}
 \label{montecarlo}
\end{figure}

In Fig.~\ref{montecarlo}, we plot the normalized SPDM $\rho^{(1)}(x, x')/\sqrt{\rho(x)\rho(x')}$ as function of $x$, 
and at fixed $x'=0$, for relatively large values of $g$.
The gray circles in the top panel are taken from the Monte Carlo data of Ref.~\cite{Minguzzi},
while the solid lines 
show the comparison of MCTDH results with the 1D variant of the 3D analytical solution for $N=2$ bosons in a harmonic trap \cite{Busch1998,Calarco}.  
We observe that the qualitative behavior of the MCTDH results is in accord with 
the analytical result as well as with the hard-core Monte Carlo calculations -- the dip in the first-order correlations located at approximately $x=l$ is consistently visible.
Note that this dip in the correlation function $\rho^{(1)}(x, x')$ corresponds to 
a peak in {\it phase fluctuations}, defined according to \cite{Marchukov}  
$
 \langle \hat {\psi^{\dagger}}(x) \hat \psi(x') \rangle = 
 \sqrt{\rho(x) \rho(x')} 
 \exp[-\frac{1}{2}\langle \hat{\delta\phi}_{xx'}^2 \rangle],
$
where $\hat {\delta \phi}_{xx'} = \hat \phi(x) - \hat \phi(x')$ is the phase difference operator and 
$\rho(x)=\langle  \hat {\psi^{\dagger}}(x) \hat \psi(x) \rangle$ 
is the mean local density. 
For a more detailed discussion of phase fluctuations and their relation to the 
first-order correlations, we refer the reader to 
Ref.~\cite{Marchukov}.

The correlation dip is due to the presence and geometry of the trap and, consequently, related to the shape of the occupied orbitals and exists even for relatively small interaction couplings.
The built-in self-consistency of the MCTDH method is crucial in order to correctly describe the correlation phenomena in trapped quantum many-body systems, because the depth and location of the correlation dip sensitively depends on the self-consistently determined orbital shape. 

We note in the top panel of Fig.~\ref{montecarlo} 
a sizable quantitative difference to the analytical solutions 
already for interaction strengths that are far below the hard-core limit
of $g\rightarrow\infty$. However, for couplings commonly realized in experiments with magnetic traps (see for concrete estimates below), the agreement between the analytical results and MCTDH is very satisfactory, see the lower panel of Fig.~\ref{montecarlo}, even for the relatively modest number of orbitals $M=10$ used in these calculations. 
The characteristic dip in the correlation function appears for any interaction strength and is correctly reproduced by the MCTDH method to good accuracy 
in its location, while the depth of the dip 
is somewhat exaggerated by MCTDH in particular 
for larger than intermediate couplings, $g\gg 1$. 

\subsection{Fourier space}
{An experimentally accessible quantity is the Fourier transform of the first-order correlation function 
\cite{coh_tan_advances}.
In Fig.~\ref{fig_Fourier}, we show the single-particle momentum density distribution, which is defined as a 
Fourier transform of the first-order correlation function in position space according to
\begin{equation}
\tilde \rho^{(1)}(k, k') = \int dx \int dx' \rho^{(1)}(x,x') \exp[i(kx- k'x')]\\
\end{equation}
for both the MCTDH result as well as the exact solution. 
The momentum density $\tilde \rho(k)$ is a Fourier transform of the spatial density $\rho(x)$ 
serves to normalize $\tilde \rho^{(1)}(k, k')$. We see that, similarly to the real space SPDM, 
MCTDH and exact result agree quite well with each other, even for the 
very large values up to $g=1000$ we have addressed in our calculations.
\begin{figure}[t]
\center
\includegraphics[width=0.6\textwidth]{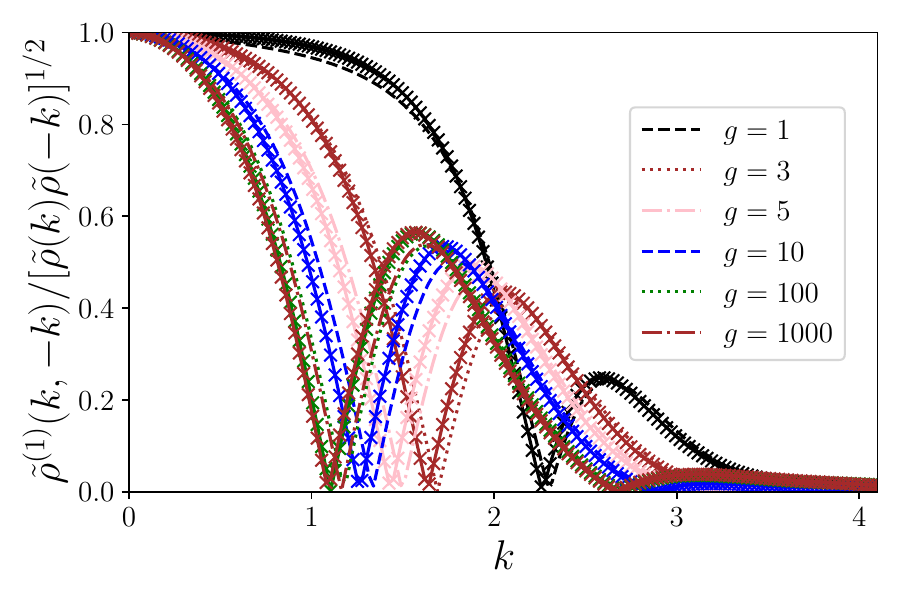}
 \includegraphics[width=0.6\textwidth]{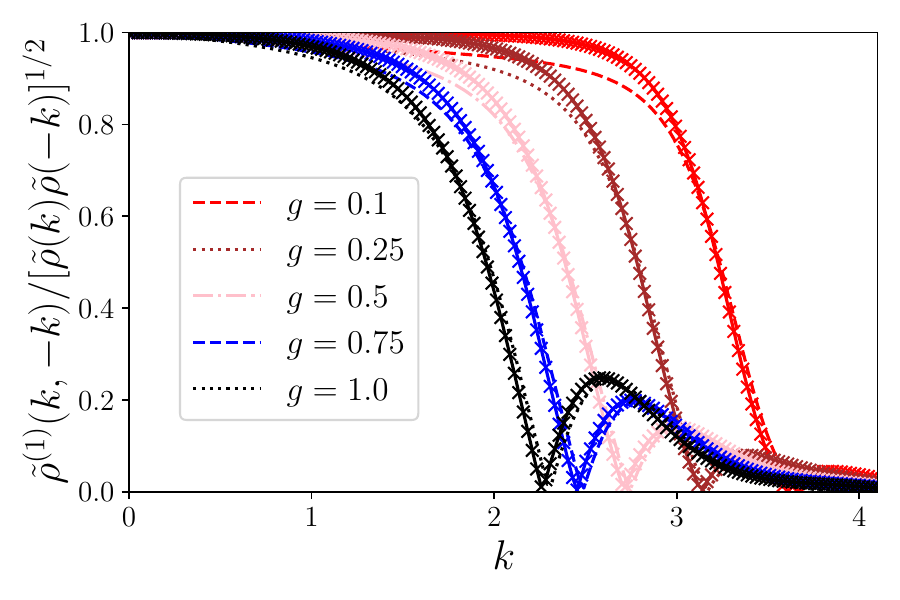}
 \caption{Normalized single-particle momentum density distribution, for various $g$ as indicated in the insets. The crosses are MCTDH results for various $g$ and $M=10$. The lines show the analytical result. }
 \label{fig_Fourier}
\end{figure}
}


\section{Fragmentation measure}  
\subsection{Definition} 
Using the SPDM, one may 
formally define an important figure of merit, the fragmentation. 
By diagonalizing the SPDM, one 
obtains its eigenfunctions, $\phi_i$, and eigenvalues, $N_i$,
which are in the many-body context referred to as 
{\it natural orbitals} and {\it occupation numbers}, respectively, 
\begin{equation}
\rho^{(1)}(x, x') = \sum_{i=0}^{M-1} N_i \phi^{\ast}_i(x') \phi_i(x).
\label{nat_orb}
\end{equation}
Here, the sum for MCTDH runs over  the finite set $i=\{0,\ldots,M-1\}$ and for the exact solution
over an infinite set $i=\{0,\ldots,\infty\}$. 

The conventional definition of fragmentation \cite{penrose1956} states that if a subset of the eigenvalues $N_i$ are ``macroscopic,''  
i.e. some of the $n_i=N_i/N$ remain finite in the large $N$ limit,
then the many-body state of the Bose gas  is a simple or fragmented BEC 
when the cardinality of this subset is one or larger than one, 
respectively \cite{leggett2003,mueller2006}. This formal definition 
of fragmentation obviously cannot be applied in our case.
We therefore {\em define} instead as the ``degree of fragmentation'  the quantity 
\begin{equation}\label{def_f} 
 \mbox{Fragmentation}  \qquad f \coloneqq\frac {N - N_0}{N},  
\end{equation}
that is, the relative occupation number of all orbitals excluding the most populated one 
(which has $i\coloneqq 0$), sorting occupation numbers $N_i$ from largest to smallest.
The fragmentation thus defined represents a shorthand for the relative number of particles residing in all orbitals outside the most occupied one (the ``condensate"). 
We note that in this definition we also take automatically into account 
the fact that the field operator truncation always has a finite number of modes $M$).

\begin{figure}[t]
\center
 \includegraphics[width=0.6\textwidth]{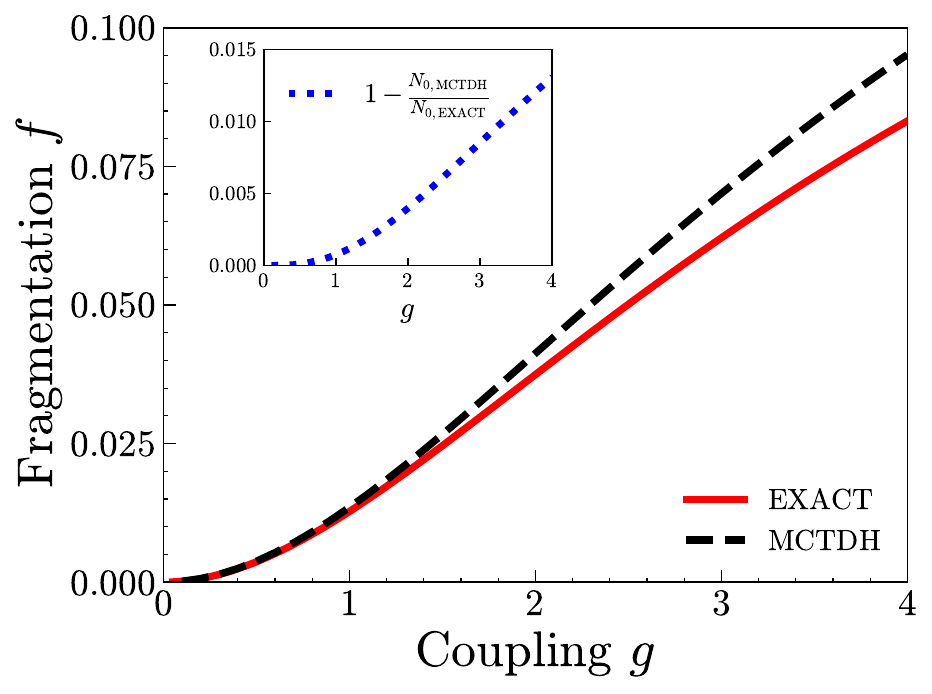}
 \caption{Fragmentation $f$ as defined in Eq.~\eqref{def_f}, obtained from the diagonalization of the analytical SPDM (red solid) and from MCTDH (black dashed) with $M=10$ orbitals, for the range $g = 0.1,\dots,4$. The inset shows the relative numerical error in the occupation number of the energetically lowest orbital.}
 \label{fig_fragm}
\end{figure}

In Fig.~\ref{fig_fragm}, we display the exact fragmentation $f$ calculated using the exact density matrix in Eq.~\eqref{exact_dm}. We obtain the exact occupation numbers
by first expressing $\rho^{(1)}(x,x')$ in 
a harmonic oscillator eigenfunctions basis of dimension $M_{\rm ho}=50$ (which proved sufficiently large) and by then diagonalizing it, evaluating the integrals via the Gauss-Hermite approximation. 
The  sizable difference when $g\gg 1$, is further illustrated in the inset, which 
shows the error in the occupation of the lowest orbital, 
$1-N_{0,{\rm MCTDH}}/N_{0,{\rm exact}}$.  Note that the fragmentation 
$f$ obtained via MCTDH is always {\em larger} than the exact value, which is in agreement 
with the observation that the former approach overestimates the correlation dip in the first-order correlations (and hence also overestimates phase fluctuations), cf.~Fig.~\ref{montecarlo}.
\begin{figure}[t]
\centering
\includegraphics[width=0.48\textwidth]{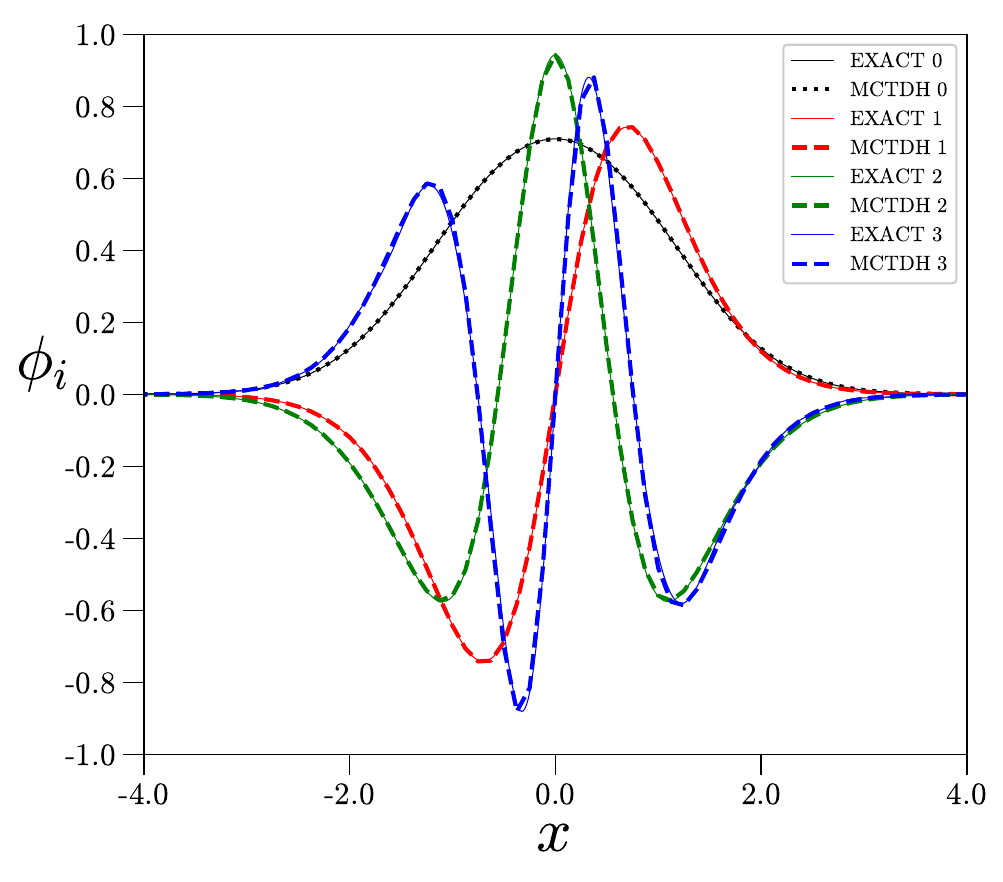}
 \hspace*{-0.5em}
 \includegraphics[width=0.48\textwidth]{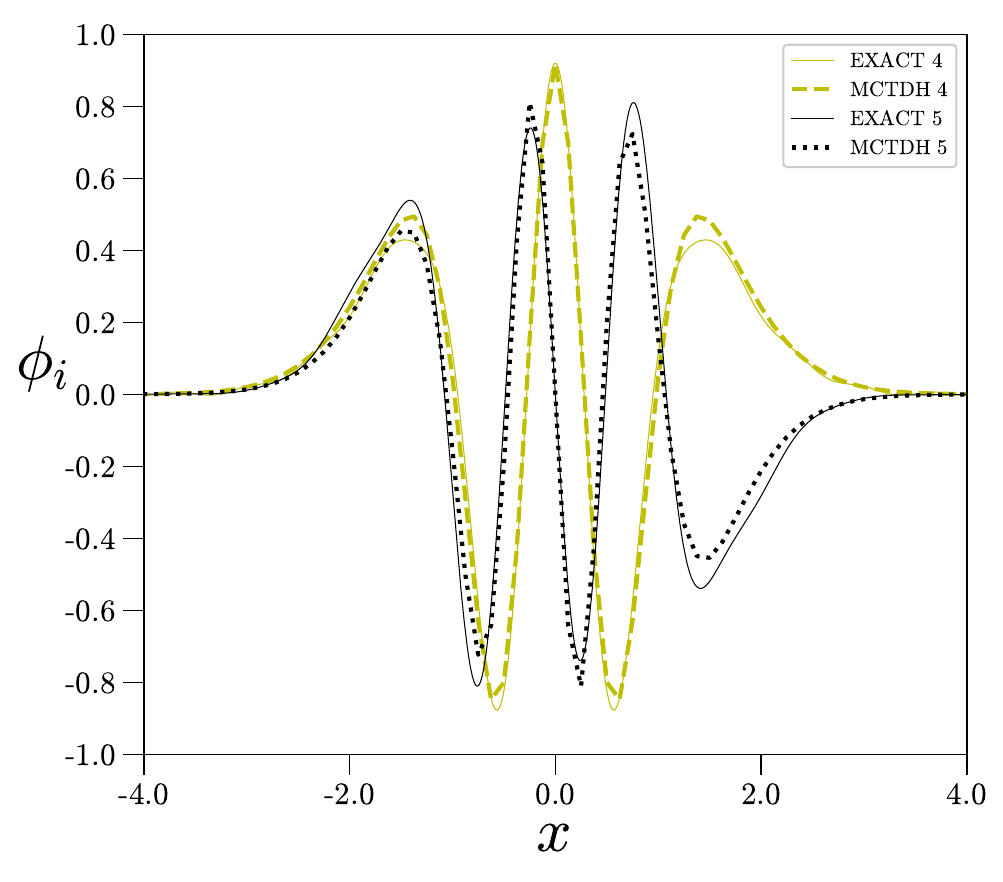}
 \includegraphics[width=0.48\textwidth]{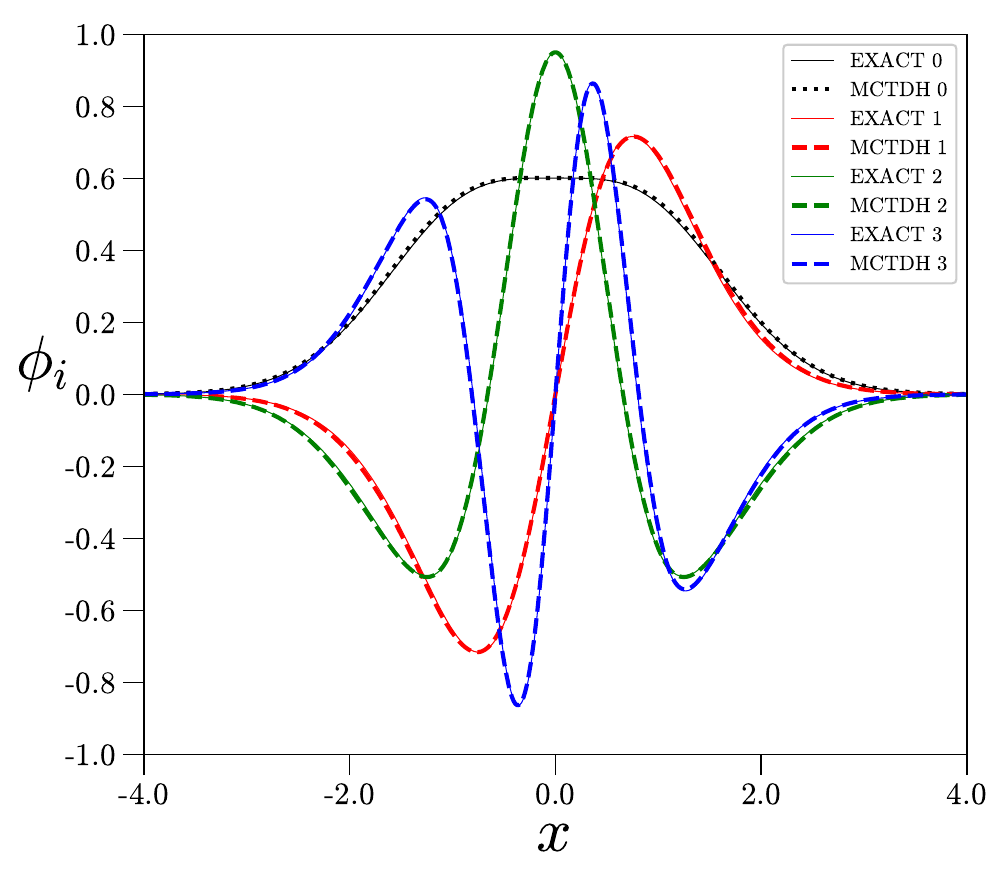}
 \hspace*{-0.5em}
 \includegraphics[width=0.48\textwidth]{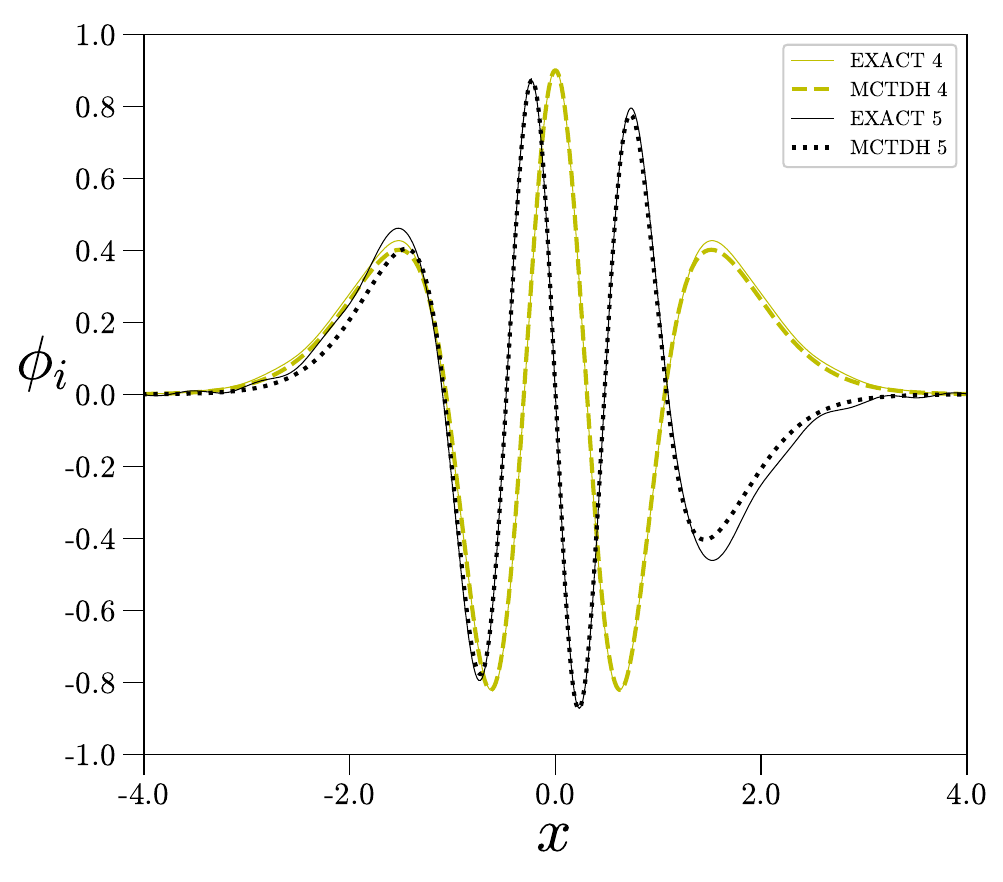} 
 \caption{The first six natural orbitals $\phi_i(x)$,  $i=0,\dots, 5$, in Eq.~\eqref{nat_orb}, 
 obtained  via MCTDH (dashed, $M=10$) and the exact results (solid), from diagonalizing the 
 SPDM in Eq.~\eqref{exact_dm} (solid).
 Left: $i=0,\dots,3$, right: $i=4,5$. Top row: $g=1$, bottom row $g=10$.}  
 \label{fig_nat_orb}
\end{figure}

\subsection{Natural orbitals}
In Fig.~\ref{fig_nat_orb}, we plot the first six  natural orbitals contained in the diagonalized SPDM 
Eq.~\eqref{nat_orb}. We conclude that sizable deviations between exact and MCTDH natural orbitals start to occur for $i=4$ and above; within the resolution of the figure, we detected no discernible deviation in the first four, that is energetically lowest, natural orbitals, $i=0,\dots,3$, the exact and MCTDH curves lying precisely on top of each other in this range. 
We also note in this context  
that the occupation numbers $N_i$ for $i>2$ are very small. For example, 
$N_3$ is about an order  of magnitude less than $N_2$, for both $g=1$ and 
$g=10$ and  for both MCTDH and exact occupation numbers \footnote{Specifically,  for $g=1$,  $N_2\simeq 2.09\times 10^{-3}$ (exact), $N_2\simeq 2.20\times 10^{-4}$ (MCTDH, $M=10$), $N_2 \simeq 2.14\times 10^{-3}$ (MCTDH, $M=33$), while 
 $N_3\simeq 4.12\times 10^{-4}$ (exact), $N_3\simeq 4.34\times 10^{-4}$ (MCTDH, $M=10$), $N_3\simeq 4.27\times 10^{-3}$ (MCTDH, $M=33$).}.  
Therefore, it is indeed the occupation number difference of the lower natural orbitals (rather than their precise shape) which explains the different fragmentation obtained by MCTDH and exact solution. As a corollary, going to much larger $M$ does not  significantly decrease the degree of fragmentation $f$-difference further. 

\section{Conclusion}   
We now illustrate the above general considerations by inserting 
concrete numbers for an experimentally realizable parameter set. In a {\it quasi-1D} Bose gas, and far away from geometric resonances \cite{olshanii1998}, we have $g = 4a_{\rm sc} l/l_\perp^2$ where $l_\perp$ is the transverse trapping length. For $^{87}$\!Rb, this implies that the coupling $g=1.96\times a_{\rm sc}[a_{\rm Rb}] 
\nu_\perp[{\rm kHz}]/\sqrt{\nu[{\rm Hz}]}$, where the background scattering length $a_{\rm Rb}=5.29$\,nm, 
$\omega_{\perp,\nu}=2\pi \nu_{\perp,\nu}$, 
and the frequencies are scaled with typical experimental values see, e.g., \cite{Betz,Fang}.   
With the background scattering length 
of $^{87}$\!Rb, and $g\sim \ord(1)$, 
the MCTDH results are in satisfactory accord with the analytical result for
quasi-1D setups accessible by magnetic trapping. 

Limits of the MCTDH approach can be explored,  
e.g., in optical lattices when one increases $g$ towards the Tonks-Girardeau regime \cite{paredes2004,kinoshita2004}. While only at a filling of two per one-dimensional tube our results can strictly be applied, 
we anticipate that also for larger $N$ 
qualitatively similar features as those  
in Fig.\,\ref{montecarlo}, and in particular the trap-induced correlation dip, 
should persist and be observable for example with (a combination of) Feshbach resonances \cite{chin2010} and higher aspect ratios. Variation of  $g$ and $N$ and measurement of, e.g., the first-order correlations which have been investigated here paves the way for a quantitative experimental assessment of the accuracy and convergence of MCTDH.
The detailed analysis of higher-order correlations \cite{schweigler2015} will then reveal further precise information on the applicability of the MCTDH method to strongly correlated systems and probe its range of applicability. 

\section*{Acknowledgments} 
OVM would like to thank D. Fedorov, A. Jensen, S. Mistakidis and A. Volosniev for useful discussions. This research was supported by the National Research Foundation of Korea, Grant No.~2014R1A2A2A01006535, Grant No. 2017R1A2A2A05001422, and Grant No.~2020R1A2C2008103.  
OVM acknowledges support by Grant No. 2015616 of 
the Binational USA-Israel Science Foundation and from the German Aeronautics and Space Administration (DLR) through Grant No. 50 WM 1957.

\bibliography{benchmark14}


\end{document}